\documentclass[twocolumn,english,prl,showpacs,aps,amsfonts,floatfix]{revtex4}

\usepackage[latin1]{inputenc}
\usepackage{amsmath}
\usepackage{graphicx}
\usepackage{amssymb}
\usepackage{amscd}
\usepackage{bm}
\usepackage{babel}

\begin{document}

\title{Multipartite Entanglement Signature of Quantum Phase Transitions}

\author{Thiago R. de Oliveira}
\email{tro@ifi.unicamp.br} 
\author{Gustavo Rigolin}
\author{Marcos C. de Oliveira} 
\author{Eduardo Miranda}

\affiliation{Instituto de Física Gleb Wataghin, 
Universidade Estadual de Campinas, CEP 13083-970, 
Campinas, S\~ao Paulo, Brazil}

\begin{abstract} We derive a general relation between the
non-analyticities of the ground state energy and those of a subclass of the 
\textit{multipartite} 
generalized global entanglement (GGE) measure defined by 
T. R. de Oliveira {\it et al.} [Phys. Rev. A \textbf{73},
010305(R) (2006)] for many-particle systems. 
We show that GGE signals both a critical point location and the order 
of a quantum phase transition (QPT). We also show that GGE allows us to study 
the relation between multipartite entanglement and QPTs, suggesting that
multipartite but not bipartite entanglement is favored at the critical 
point.   
Finally, using GGE we were able, at a second order QPT, to define a diverging 
entanglement length (EL) in terms of the usual correlation length. 
We exemplify this with the XY spin-1/2 chain and show that the EL 
is half the correlation length.

\end{abstract}

\maketitle

Quantum Phase Transitions (QPTs) occur at zero temperature and are
characterized by non-analytical changes in the physical properties of
the ground state of a many-body system governed by the variation of a
parameter $\lambda$ of the system's Hamiltonian $H(\lambda)$. These
changes are driven solely by quantum fluctuations and are usually
characterized by the appearance of a non-zero order parameter
\cite{livro-do-sachdev}.  Since QPTs occur at $T=0$, the emerging
correlations have a purely quantum origin. Therefore, it is reasonable to 
conjecture that
entanglement is a crucial ingredient for the occurrence of QPTs
(e.g. Refs. \cite{nature,nielsen,latorre1,thiago1} and references
therein). If this is true, then the QPT would imprint its signature on
the behavior of an entanglement measure.  
Under a set of reasonable general assumptions, Wu \textit{et al.} 
\cite{sarandy1} have demonstrated that a discontinuity in a \textit{bipartite}
entanglement measure (concurrence \cite{wootters} and negativity
\cite{werner}) is a necessary and sufficient indicator of a 
first order quantum phase transition (1QPT),
the latter being characterized by a discontinuity in the first derivative
of the ground state energy. Furthermore, they have shown that 
a discontinuity or a divergence in the first derivative of the
same measure (assuming it is continuous) is a
necessary and sufficient indicator of a second order QPT (2QPT), which 
is characterized by a discontinuity or
a divergence of the second derivative of the ground state energy. 
Nevertheless, most of the
models of 2QPTs considered so far did not present any \textit{long-range} 
bipartite entanglement at the critical point, even though the correlation 
length diverges. Moreover, contrary to expectations, most of the
measures discussed in the literature are, to the best of our knowledge, not
maximal at the critical point, the exceptions being the one-site von
Neumann entropy of the Ising chain \cite{nielsen}, the localizable
entanglement of a finite Ising chain with 14 sites \cite{LE}, and
some classes of the Generalized Global Entanglement (GGE) for the
Ising chain \cite{thiago1}.

In this Letter we firstly extend Wu \textit{et al.} \cite{sarandy1} results to
a \textit{multipartite} entanglement  (ME) measure 
\cite{foot-sarandy2,sarandy2}, the GGE introduced in
Refs. \cite{thiago1,thiago2}, and discuss how non-analyticities in the
energy are signaled by the GGE. Secondly, we define an entanglement 
length (EL) for an arbitrary
collection of two-level systems. In the case of a symmetry-breaking
2QPT, this EL diverges at the critical point and is
simply related to the correlation length. This result indicates 
that ME is most favored at that point, contrary to bipartite
entanglement \cite{nature,nielsen,tommaso}. We consider the consequences
of this result for specific spin-1/2 models presenting 1QPT or 2QPT.

In particular, for the 2QPT of the one-dimensional transverse field XY
model we obtain all the relevant critical exponents, with the
EL defined in terms of correlation functions (CFs)
appearing in the GGE. We also show in this specific case that the GGE
is maximal at the critical point, thus signaling the QPT, as three of us had
already observed in the Ising case \cite{thiago1}. This last result,
together with a diverging ME length at the critical point, reinforces 
that ME plays a significant role in QPTs.

Following Ref. \cite{sarandy1}, a discontinuity in (discontinuity
in or divergence of the first derivative of) the concurrence or negativity
is both necessary and sufficient to signal a 1QPT (2QPT) for systems of 
distinguishable particles governed by up to two-body Hamiltonians. The energy
per particle ($\varepsilon$) derivatives depend on the
two-particle density matrix elements as \cite{sarandy1} 
\begin{eqnarray}
\partial_{\lambda}\varepsilon&=&(1/N)\sum_{ij}Tr\left[\left(\partial_{\lambda}U\left(i,j\right)\right)\rho_{ij}\right],\\
\partial_{\lambda}^{2}\varepsilon & = & (1/N)\sum_{ij}\left\{ Tr\left[\left(\partial_{\lambda}^{2}U\left(i,j\right)\right)\rho_{ij}\right]\right.\nonumber \\
 &  &+ \left.Tr\left[\left(\partial_{\lambda}U\left(i,j\right)\right)\partial_{\lambda}\rho_{ij}\right]\right\} \label{der 2 E},\end{eqnarray}
where $\rho_{ij}$ is the reduced two-particle density operator and 
$U\left(i,j\right)$ includes all the single and two-body terms of the 
Hamiltonian associated with particles $i$ and $j$.
Now, assuming that $U\left(i,j\right)$ is a smooth function of the
Hamiltonian parameters and that $\rho_{ij}$ is finite at the critical
point, {\it the origin of the discontinuity in the energy (discontinuity
in or divergence of the first derivative of the energy) is 
the fact that one or more of the elements of $\rho_{ij}$ 
($\partial_{\lambda}\rho_{ij}$)
are discontinuous (divergent) at the transition point $\lambda=\lambda_c$} 
\cite{sarandy1}. 
Since the concurrence and the negativity are both linear functions
of the elements of $\rho_{ij}$ it turns out that a discontinuity/divergence
in one of them (in the derivative of one of them) implies a 
discontinuity/divergence
of the energy (in the derivative of the energy) and vice-versa \cite{sarandy1}.
A natural question then arises: 
Does a ME measure show the same feature? In what 
follows, we give an explicit affirmative answer to this question
\cite{foot-sarandy2}.  

In \cite{thiago1} three of us introduced two new quantities, both of which can 
be seen 
as generalizations of the Meyer-Wallach \cite{meyer} Global Entanglement,
originally defined for a system of $N$ parties (particles). The first one is 
the average linear entropy of all $N_{1}<N$ particles, where we assume a fixed 
``distance'' between the $N_{1}$ particles. 
The second quantity is an average over all possible distances/configurations 
in which the $N_1$ particles can be arranged \cite{thiago1,thiago2}. For 
$N_{1}=1$ both quantities are the same ($G(1)=E_{G}^{(1)}$)
and we recover the Meyer-Wallach measure. The 
first non-trivial case appears when $N\geq 4$ and we pick two particles 
($N_1=2$) labeled by $i$ and $j$. 
Now for a density matrix $\rho_{j,j+n}$ of dimension $d$ we have $G\left(2,n\right)=\frac{d}{d-1}\left[1-(1/(N-n))\sum_{j=1}^{N-n}Tr\left(\rho_{j,j+n}^{2}\right)\right]$, which is the mean linear entropy of all pairs of 
particles $n=|i-j|$ sites apart, i.e., the mean entanglement between these pairs and the remaining $N-2$ particles. Averaging over 
all possible distances $1\leq n<N$, $E_{G}^{(2)}$ $=$ 
$\frac{2}{N(N-1)} \sum_{n=1}^{N-1}(N-n){G\left(2,n\right)}$.  In order to simplify the notation (with no loss
of generality), from now on we will work with the linear entropy of a single 
pair of particles $n$ sites apart, which we call $\mathcal{G}(2,n)$. Note that 
in this notation $G(2,n)=\overline{\mathcal{G}(2,n)}$, being the average of $\mathcal{G}(2,n)$ over all particles $n$ sites apart. For a translationally
symmetric system $G(2,n) = \mathcal{G}(2,n)$.  

Considering $\mathcal{G}\left(2,n\right)$ as a function of the tuning parameter 
$\lambda$ we can write it and its derivative in terms of the $lm$ elements of 
$\rho_{ij}$ ($\left[\rho_{j,j+n}\right]_{lm}$) as
\begin{eqnarray}
\mathcal{G}\left(2,n\right)&=&\frac{d}{d-1}\left[1-\sum_{l,m=1}^{d^2}\left|\left[\rho_{j,j+n}\right]_{lm}\right|^{2}\right], \label{G} \\
\partial_{\lambda}\mathcal{G}\left(2,n\right)&=&\frac{2d}{1-d}\!\sum_{l,m=1}^{d^2}\!\!\left|\left[\rho_{j,j+n}\right]_{lm}\right|\partial_{\lambda}\left|\left[\rho_{j,j+n}\right]_{lm}\right|. \label{delG} \end{eqnarray}

{\it Therefore, since a discontinuity in one or more 
$\left[\rho_{j,j+n}\right]_{lm}$ signals a 1QPT, a discontinuity in 
$\mathcal{G}\left(2,n\right)$ also signals a 1QPT. 
If $\mathcal{G}\left(2,n\right)$ is continuous and 
$\partial_{\lambda} \mathcal{G}\left(2,n\right)$ shows a discontinuity
or divergence, it signals a 2QPT}. 
In this sense $\mathcal{G}\left(2,n\right)$
is at least as good as the concurrence/negativity to signal a QPT. Note that
the previous result is valid only if the discontinuous/divergent quantities do
not accidentally all vanish or cancel with other terms in Eqs. (\ref{G}) and 
(\ref{delG}) (assumptions (b) and (c) in 
Ref. \cite{sarandy1}). An added bonus of our approach, however, is
that we do not need a further assumption, as in 
Ref. \cite{sarandy1}, related to the 
artificial/accidental divergences due to the maximization/minimization processes 
appearing in the definitions of the concurrence and the negativity. 
Moreover, $\mathcal{G}\left(2,n\right)$ is richer than the concurrence/negativity for signaling and classifying
the order of a QPT,
since it can be employed for the derivation of an EL, as we now demonstrate.

We particularize our discussion to two-level (qubit) systems \cite{livro-do-sachdev}. 
In this case $\mathcal{G}\left(2,n\right)$ is written as 
\begin{equation}
\mathcal{G}\left(2,n\right)=\frac{4}{3}\left[1-\frac{1}{4}\sum_{\alpha,\beta=0}^3\langle\sigma_{j}^{\alpha}\sigma_{j+n}^{\beta}\rangle^{2}\right],\label{gcorr}
\end{equation} 
where $\sigma_i^\alpha,\,\alpha=1,2,3$, are the Pauli operators and $\sigma_i^0$ is the identity. Thus, \textit{as any measure dependent only on the two-particle reduced 
density matrix}, $\mathcal{G}\left(2,n\right)$ is \textit{completely} determined by one- and 
two-point CFs. 
Whenever the system undergoes a second-order
symmetry-breaking QPT, it will be reflected in one or more CFs and hence in the behavior of 
$\mathcal{G}(2,n)$. If the dominant (less rapidly decaying) CF 
decays with a power law ($\langle\sigma_{i}^{\alpha_{0}}\sigma_{j}^{\beta_{0}}\rangle\sim n^{-\eta}$)
at the critical point (implying a diverging correlation length) and exponentially in its vicinity
($\langle\sigma_{i}^{\alpha_{0}}\sigma_{j}^{\beta_{0}}\rangle\sim e^{-n/\xi_{C}}$), so will
$\mathcal{G}\left(2,n\right)$ increase. For large
$n$,
%
$\mathcal{G}\left(2,n\right)\approx \mathcal{G}\left(2,\infty\right)-\langle\sigma_{i}
^{\alpha_{0}}\sigma_{j}^{\beta_{0}}\rangle^{2}/3.$ Hence, close to the 
critical point
$\mathcal{G}\left(2,n\right)\approx \mathcal{G}\left(2,\infty\right)- C e^{-2n/\xi_{C}}$, 
where $C$ is a constant, and
$\mathcal{G}\left(2,n\right)$ increases exponentially fast,
saturating for $n \gg \xi_{C}/2$. We can then define an
EL that is proportional to the correlation length,
$\xi_{E}=\xi_{C}/2$. The EL also diverges at
the critical point with the same exponent as $\xi_C$, such that
$\xi_{E}\sim|\lambda-\lambda_{c}|^{-\nu}$.
At $\lambda=\lambda_c$, for large $n$,
$
\mathcal{G}\left(2,n\right)\approx 
\mathcal{G}\left(2,\infty\right)- C^{\prime}n^{-2\eta},
$
where $C^{\prime}$ is a constant. 
$\mathcal{G}\left(2,n\right)$ now increases as a power law
with a power that is twice the CF exponent. 
Thus,  $\mathcal{G}\left(2,n\right)$ inherits all the universal properties
of the CFs. Moreover, due to the $\mathcal{G}\left(2,n\right)$ scaling with $n$, at the critical point the entanglement
is more \textit{distributed} in the system (any two spins are entangled
with the rest of the chain) than away from it, 
indicating ME \cite{nielsen} prevails at the critical point.
We emphasize that this result is quite general, applying to any collection of
two-level systems with a 
second-order symmetry-breaking QPT. 
Next, we particularize to two specific cases in order to illustrate our general results.

For an arbitrary spin-1/2 model presenting a 1QPT, at least one of the
CFs is discontinuous at the transition point. Thus, it is intuitive that
$\mathcal{G}\left(2,n\right)$ is also discontinuous. A simple example
is the {\it frustrated two-leg spin-1/2 ladder} discussed in
Refs. \cite{sarandy1,Chattopadhyay}, where all but the $\langle
\sigma_i^\alpha\sigma_j^\alpha\rangle$, $\alpha = x, y, z$, and
$\langle \sigma_i^z\rangle$ expectation values vanish. The latter are
discontinuous at the transition point but constant otherwise.  The
transition is clearly of first order and $\mathcal{G}\left(2,n\right)$
is able to signal it.

{\it The one-dimensional
XY model in a transverse magnetic field}
is described by the following Hamiltonian\begin{equation}
H\!\!=-\sum_{i=1}^{N}\!\!\left\{ \frac{J}{2}\left[\left(1+\gamma\right)\sigma_{i}^{x}\sigma_{i+1}^{x}+\left(1-\gamma\right)\sigma_{i}^{y}\sigma_{i+1}^{y}\right]+h \sigma_{i}^{z}\right\}, \end{equation}
 where $N$ is the total number of spins (sites) and $\gamma>0$ is the anisotropy.
This Hamiltonian is symmetric under a global $\pi$ rotation about
the z axis ($\sigma_{i}^{x(y)}\rightarrow-\sigma_{i}^{x(y)}$), implying a zero
 magnetization in the $x$ or $y$ directions 
($\langle \sigma_{i}^{x(y)}\rangle=0$).
However, as the magnetic field $h$ is decreased (or $J$ increased)
this symmetry is spontaneously broken at $\lambda=J/h=1$ (in the
thermodynamical limit) and a doubly-degenerate ground state with finite
magnetization ($\pm M$) in the $x$ direction develops, characterizing a 
ferromagnetic phase. It is possible then to define a symmetric ground state (with $\langle \sigma_{i}^{x}\rangle=0$) as a
superposition of these two degenerate ones. 
These states are of no use in practice, however, as they do not exist 
in real macroscopic objects undergoing a phase transition (``clustering property'').
We call non-symmetric or broken-symmetry states the ones in which there is a finite magnetization 
($\langle S_{i}^{x} = \sigma_i^x/2 \rangle=\pm M$).  
Note that at the paramagnetic phase there is no such distinction.
By further decreasing the magnetic field, a second
phase transition occurs at $\gamma^{2}+h^{2}=1$. In this 
``third'' phase the approach of the CFs to their 
saturation values is not monotonic but oscillatory \cite{McCoy1}. 
We should also say that this model
reduces to the Ising model for $\gamma=1$, where only the
first critical point exists, and to the XX model as $\gamma\rightarrow 0$.
However, the XX model belongs to a different universality class and we consider here
only $0<\gamma \le 1$ \cite{McCoy1}.

The XY model can be solved exactly and all the CFs
are known \cite{McCoy1}. To calculate
$\mathcal{G}\left(2,n\right)$ all we need is 
the one and two-point CFs (See Eq. (\ref{gcorr})). 
Due to the translational invariance of the model $\rho_{ij}$ depends only on 
the distance $n=|i-j|$ between the spins and
$\langle\sigma_{i}^{\alpha}\sigma_{j}^{\beta}\rangle
=\langle\sigma_{j}^{\alpha}\sigma_{j+n}^{\beta}\rangle
=p_n^{\alpha\beta}$. 
%
Remembering that $\rho_{ij}$ is Hermitian and has a unitary trace we are left 
with nine independent elements of $\rho_{ij}$, which may be functions of 
at most nine one- and two-point CFs. 
%
This number can be further reduced by the symmetries
of the problem. The global symmetry under a $\pi$ rotation
about the $z$ axis yields
$\langle\sigma^{x(y)}\rangle=\langle\sigma_{i}^{x}\sigma_{i}^{y}\rangle
=\langle\sigma_{i}^{x}\sigma_{i}^{z}\rangle
=\langle\sigma_{i}^{y}\sigma_{i}^{z}\rangle=0$
in the paramagnetic phase ($\lambda\leq1$). We end up with four elements: 
$\langle\sigma^{z}\rangle$ and 
$\langle\sigma_{i}^{\alpha}\sigma_{i}^{\alpha}\rangle$, $\alpha=x,y,z$.
In the ferromagnetic phase ($\lambda>1$) this no longer holds 
since the Hamiltonian symmetry is not preserved by the ground
state and we need to evaluate the nine one and two-point
CFs. The four CFs appearing in the paramagnetic
phase and $\langle\sigma_{i}^{x(y)}\rangle$ plus the three off-diagonal 
two-point ones 
were calculated in Ref. \cite{McCoy1}.
%
The first two $p_{n}^{yz}=p_{n}^{xy}=0$
for all values of $\gamma$ and $\lambda$ \cite{McCoy1}. 
The last off-diagonal CF ($p_{n}^{xz}$) was obtained
exactly 
in terms of complex integrals whose calculation is cumbersome. 
However, we were able to obtain excellent 
bounds for it by imposing the positivity of the eigenvalues of $\rho_{ij}$  
\cite{thiago2}.
%

\begin{figure}[!ht]
\includegraphics[width=2.75in]{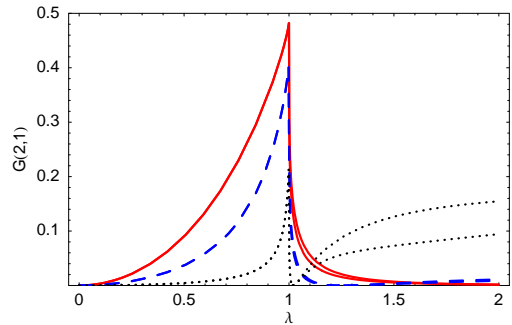}
\caption{\label{Graf G21} (Color online) Upper and lower bounds of 
$\mathcal{G}(2,1)$ for the XY chain for three values of the anisotropy: 
$\gamma=1$ (red/solid), $0.6$ (blue/dashed), and $0.2$ (black/dotted).}
\end{figure}

\begin{figure}[!ht]
\includegraphics[width=2.75in]{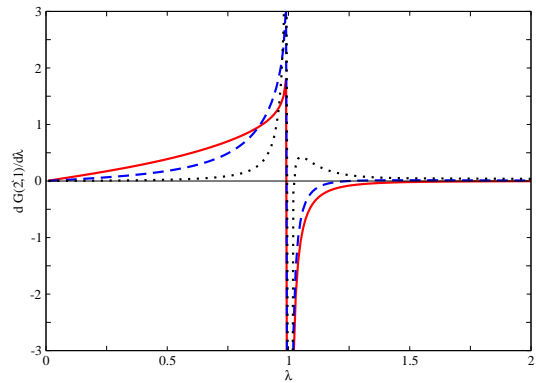}
\caption{\label{Graf Der G21}(Color online) Derivative of the lower bound of 
$\mathcal{G}(2,1)$ for three values of anisotropy: $\gamma=1$ (red/solid), 
$0.6$ (blue/dashed), $0.2$ (black/dotted). The second phase transition is also 
imprinted for the $\gamma=0.2$ as the curve crosses the abcissa at 
$\lambda=1/\sqrt{1-\gamma^2}$.}
\end{figure}

With all the necessary CFs in hand  $\mathcal{G}\left(2,n\right)$ for the 
XY model reads
$
\mathcal{G}(2,n)$ $=$ 
$1$$-$$\frac{1}{3}$$[2\langle\sigma_{j}^{x}\rangle^{2}$
$+$$2\langle\sigma_{j}^{z}\rangle^{2}
+2\langle\sigma_{j}^{x}\sigma_{j+n}^{z}\rangle^{2}+
\langle\sigma_{j}^{x}\sigma_{j+n}^{x}\rangle^{2}
+\langle\sigma_{j}^{y}\sigma_{j+n}^{y}\rangle^{2}
+\langle\sigma_{j}^{z}\sigma_{j+n}^{z}\rangle^{2}].
$
In Fig.~\ref{Graf G21} we plot the lower and upper bounds for
$\mathcal{G}\left(2,1\right)$ (by using the upper and lower bounds of $p_{n}^{xz}$, respectively) as a function
of $\lambda$  and for a few $\gamma$'s. We first note that it is maximal at the 
critical point for any anisotropy (this is true throughout the interval
$0<\gamma\le 1$). Secondly, the bounds obtained are very tight and can 
barely be distinguished for some anisotropies. 
Only in the ferromagnetic phase and for $\gamma\rightarrow 0$ do the bounds become distinguishable. 
The derivative of $\mathcal{G}\left(2,1\right)$ with respect to
$\lambda$  is depicted in Fig.~\ref{Graf Der G21} exhibiting, 
as expected, a divergence at the critical point.
%
%
%
%

\begin{figure}[!ht]
\includegraphics[width=2.75in]{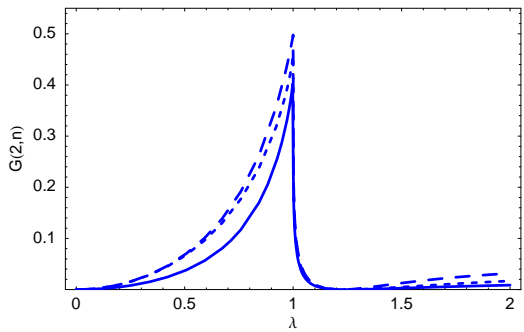}
\caption{\label{Graf G212227} (Color online) Lower bound of 
$\mathcal{G}\left(2,n\right)$ for $\gamma=0.6$ and for three values of 
$n$: $n=1$ (solid), 2 (dashed), and 7 (long-dashed).}
\end{figure}

Now we analyse how $\mathcal{G}(2,n)$ approaches its asymptotic value. It can be seen
in Fig. \ref{Graf G212227} that 
$\mathcal{G}\left(2,n\right)$ is an increasing function of the distance $n$. 
To study this behavior analytically we make use of the asymptotic form of the CFs of the XY model: for $\lambda<1$  \cite{McCoy1}, $
\langle\sigma_{j}^{x}\sigma_{j+n}^{x}\rangle\sim n^{-1/2}\lambda_{2}^{n}$,
 $
\langle\sigma_{j}^{y}\sigma_{j+n}^{y}\rangle\sim n^{-3/2}\lambda_{2}^{n}$, and 
$
\langle\sigma_{j}^{z}\sigma_{j+n}^{z}\rangle\sim\langle\sigma^{z}\rangle^{2}-n^{-2}\lambda_{2}^{2n}$,
 with $
\lambda_{2}=(1/\lambda-\sqrt{1/\lambda^{2}-\left(1-\gamma²\right)})/(1-\gamma)$,
 while at the critical point $
\langle\sigma_{j}^{x}\sigma_{j+n}^{x}\rangle\sim n^{-1/4}$,
 $\langle\sigma_{j}^{y}\sigma_{j+n}^{y}\rangle\sim n^{-9/4}$, and
 $
\langle\sigma_{j}^{z}\sigma_{j+n}^{z}\rangle\sim\langle\sigma^{z}\rangle^{2}-n^{-2}$.
We can see from these expressions that, for large values of
$n$, the dominant correlation is, as expected, in the $x$
direction. Thus, for large $n$ we can write 
$
\mathcal{G}\left(2,n\right)\sim \mathcal{G}\left(2,\infty\right)-\langle\sigma_{j}^{x}\sigma_{j+n}^{x}\rangle^{2}/3,
$
such that
\begin{eqnarray}
\mathcal{G}\left(2,n\right)&\sim& \mathcal{G}\left(2,\infty\right)- C n^{-1}\lambda_{2}^{2n},\,
 \lambda<\lambda_c,\\
\mathcal{G}\left(2,n\right)&\sim& \mathcal{G}\left(2,\infty\right)- C^{\prime}n^{-1/2},\,\,\,\,\, \lambda=\lambda_c.\end{eqnarray}
From these expressions, we see explicitly that,
at the critical point, the entanglement between two spins $n$ sites
apart increases as a power law of their distance, whereas away from
the critical point it increases exponentially and saturates very fast.
For the XY model the EL defined before reads
$\xi_{E}=\frac{\gamma}{2\left(1-\lambda\right)}$, where we have
used that $\lambda_{2}\approx1+(\lambda-1)/\gamma$ near the critical
point. Note that $\xi_{E}$ diverges at the critical point as expected
and that the ratio between $\xi_{E}$ and the correlation length
$\xi_{C}$ is fixed: $\xi_{E}/\xi_{C}=1/2$. 
Thus at the critical point the entanglement in the XY model is 
more distributed in the
chain, as already indicated by the block entanglement \cite{latorre1}.

In conclusion, we related the non-analytic properties of the ground
state energy to the non-analyticities of $\mathcal{G}(2,n)$ for an
arbitrary many-particle system. Thus, $\mathcal{G}(2,n)$ is able to
signal both the quantum phase transition (QPT) points and the order of the
transtition. $\mathcal{G}(2,n)$ is a multipartite entanglement (ME) measure
which, for many reasons \cite{thiago2}, is operationally good. Since
no maximization/minimization process is needed for its calculation, no
accidental discontinuities or divergences will occur (in contrast to
the concurrence or the negativity). Moreover, for two-level systems
$\mathcal{G}(2,n)$ is simply related to one- and two-point
correlation functions (CFs). Therefore, for those systems undergoing a second
order QPT it is possible to
define a critical exponent and an entanglement length which is half
the more familiar correlation length. We have exemplified those
results with an explicit calculation for the XY transverse field
spin-1/2 chain. This result adds strength to the conjecture by
T. J. Osborne and M. A. Nielsen \cite{nielsen} that at the critical point
bipartite entanglement (as given by the concurrence/negativity) is not
maximal due to entanglement sharing, since all the parties involved
are entangled as the entanglement length diverges. In fact, what
should be maximal and favored is the \textit{multipartite} entanglement, as we
have plenty demonstrated. It is worth mentioning that any knowledge
of the behavior of ME can only be achieved via the generalized global 
entanglement (GGE) and not by any CF alone.
We expect that these findings will
contribute to the understanding of the relevance of entanglement,
specially ME, in QPTs.

\textit{Note:} After this work was completed we became aware of an
independent derivation of the entanglement length for the XY model in
terms of the two-site \textit{von Neumann entropy} \cite{chen}. We
point out, however, that the relatively simple form of
$\mathcal{G}(2,n)$, as given by the one and two-point CFs, allows it
to be employed for the determination of the order of the QPT as well
as for the derivation of an entanglement length for an arbitrary
two-level system undergoing a second order QPT.

We thank A.O. Caldeira for interesting discussions and 
acknowledge support from CNPq and FAPESP. 


\begin{thebibliography}{10}
\bibitem{foot-sarandy2} Note that via the Density Functional Theory \cite{sarandy2}
it is possible to arrive at similar results.
\end{thebibliography}
\end{document}